\documentclass[%
preprint,
superscriptaddress,
 amsmath,amssymb,
 aps,
]{revtex4-2}

\usepackage{graphicx}
\usepackage{dcolumn}
\usepackage{bm}
\usepackage{hyperref}
\hypersetup{
    colorlinks=true,
    urlcolor= blue,
    citecolor=blue,
linkcolor= blue}

\begin{document}

\title {Challenges to magnetic doping of thin films of the Dirac semimetal {Cd$_3$As$_2$}}
\author{Run Xiao}
\affiliation{Department of Physics, The Pennsylvania State University, University Park, PA 16802, USA}
\author{Jacob T Held}
\affiliation{Department of Chemical Engineering and Materials Science, University of Minnesota, Minneapolis, MN 55455, USA}
\author{Jeffrey Rable}
\affiliation{Department of Physics, The Pennsylvania State University, University Park, PA 16802, USA}
\author{Supriya Ghosh}
\affiliation{Department of Chemical Engineering and Materials Science, University of Minnesota, Minneapolis, MN 55455, USA}
\author{Ke Wang}
\affiliation{Materials Research Institute, The Pennsylvania State University, University Park PA 16802}
\author{K. Andre Mkhoyan}
\affiliation{Department of Chemical Engineering and Materials Science, University of Minnesota, Minneapolis, MN 55455, USA}
\author{Nitin Samarth}%
\thanks{Corresponding author: nsamarth@psu.edu}
\affiliation{Department of Physics, The Pennsylvania State University, University Park, PA 16802, USA}

\date{\today}

\begin{abstract}
Magnetic doping of topological quantum materials provides an attractive route for studying the effects of time-reversal symmetry breaking. Thus motivated, we explore the introduction of the transition metal Mn into thin films of the Dirac semimetal Cd$_3$As$_2$ during growth by molecular beam epitaxy. Scanning transmission electron microscopy measurements show the formation of a Mn-rich phase at the top surface of Mn-doped Cd$_3$As$_2$ thin films grown using both uniform doping and delta doping. This suggests that Mn acts as a surfactant during epitaxial growth of Cd$_3$As$_2$, resulting in phase separation. Magnetometry measurements of such samples indicate a ferromagnetic phase with out-of-plane magnetic anisotropy. Electrical magneto-transport measurements of these films as a function of temperature, magnetic field, and chemical potential reveal a lower carrier density and higher electron mobility compared to pristine Cd$_3$As$_2$ films grown under similar conditions. This suggests that the surfactant effect might also serve to getter impurities. We observe robust quantum transport (Shubnikov-de Haas oscillations and an incipient integer quantum Hall effect) in very thin (7 nm) Cd$_3$As$_2$ films despite being in direct contact with a structurally disordered surface ferromagnetic overlayer.
\end{abstract}
\maketitle
\section{Introduction}

Magnetic doping of quantum materials such as semiconductors and topological insulators (TIs) is a well established route to the discovery of interesting emergent quantum phenomena.\cite{MacDonald_NM,liu2008quantum,yu2010quantized,xu2012hedgehog,chang2013thin} Magnetically-doped semiconductors such as (Ga,Mn)As have provided an important platform for proof-of-concept spintronic devices \cite{dietl2014dilute} while magnetic doping of certain TIs has led to the discovery of the quantum anomalous Hall effect \cite{chang2013experimental,kandala2015giant,chang2015high}. In the latter case, the key physics is driven by the breaking of time-reversal symmetry by ferromagnetic order induced by the exchange interaction between transition metal dopants (Cr, V, Mn, and Fe) and the extended band states of the TI in the (Bi,Sb)$_2$(Te,Se)$_3$ family. 

The success of transition metal doping of semiconductors and TIs provides a strong motivation for exploring similar magnetic doping of other topological materials such as topological Dirac semimetals \cite{liu2014discovery,xu2015observation,liu2014stable,neupane2014observation,yi2014evidence,armitage2018weyl}. Topological Dirac semimetals host three-dimensional Dirac fermions and can be identified as a parent phase of other topological phases, for instance, Weyl semimetals \cite{weyl1929elektron,burkov2011weyl}. Introducing magnetic dopants into a Dirac semimetal to break the time-reversal symmetry could lead to the degenerate Dirac fermions separating into two (or four) Weyl fermions and a Weyl semimetal phase \cite{baidya2020first}.

In the last decade, Cd$_3$As$_2$ has been theoretically predicted \cite{wang2013three} and experimentally demonstrated \cite{liu2014stable,neupane2014observation,yi2014evidence}as a Dirac semimetal. More recently, high-quality Cd$_3$As$_2$ thin films have been successfully grown using molecular beam epitaxy (MBE) \cite{schumann2016molecular,uchida2017quantum,liu2015gate,yanez2021spin}  and exhibit the quantum Hall effect \cite{schumann2018observation,uchida2017quantum,Galletti_PhysRevB.99.201401}. The past success of MBE growth of magnetically-doped semiconductors provides a strong motivation to explore magnetic doping of MBE-grown Cd$_3$As$_2$ thin films as a route toward breaking time-reversal symmetry \cite{baidya2020first} for realizing novel topological phases such as magnetic Weyl semimetals \cite{liu2019magnetic,morali2019fermi,belopolski2019discovery} and monopole superconductors \cite{li2018topological}. We are aware of only two published reports on magnetic doping of Cd$_3$As$_2$ thin films using Cr \cite{liu2018cr} and Mn \cite{Wang_2020}. These studies principally relied on electrical magneto-resistance measurements to draw conclusions about the effect of magnetic dopants on the Dirac semimetal states and assumed that the magnetic dopants were homogeneously distributed throughout the Cd$_3$As$_2$ film.    

In this paper, we describe our attempts to use MBE to dope Mn into the Dirac semimetal Cd$_3$As$_2$. We present a systematic structural and magnetic study of heterostructures using high-angle annular dark-field scanning transmission electron microscopy (HAADF-STEM), energy dispersive X-ray spectroscopy (EDX), atomic force microscopy (AFM), high-resolution x-ray diffraction (XRD), superconducting quantum interference device (SQUID) magnetometry, and electrical transport. We find that instead of being incorporated into the Cd$_3$As$_2$ lattice, the Mn dopants form a Mn-rich layer on the top of the Cd$_3$As$_2$ layer. The Mn-rich layer shows insulating behavior and lowers the carrier density in the Cd$_3$As$_2$ layer underneath. Remarkably, even though the Mn-rich phase/Cd$_3$As$_2$ heterostructure exhibits ferromagnetism at room temperature with out-of-plane anisotropy, the samples show pronounced quantum oscillations and an incipient integer quantum Hall effect at low temperature. 

\section{Results and discussion}

 The Mn-doped Cd$_3$As$_2$ thin films were grown by MBE in a Veeco EPI 930 chamber. We used epi-ready miscut semi-insulating GaAs (111)B substrates ($1^{\circ}$ toward $(2\bar{1}\bar{1})$).  Elemental source materials were evaporated from standard effusion cells containing As (99.999995\%), Ga(99.99999\%), Sb(99.9999\%), Cd (99.9999\%), and Mn (99.9998\%). The epi-ready GaAs substrates were first annealed inside the MBE chamber to flash off the native oxide at a thermocouple temperature of 720 $^{\circ}$C (the actual temperature is likely 580 $^{\circ}$C). Then, we deposited a thin ($\sim 2 $nm)  GaAs layer at the same substrate temperature to smooth the surface. Subsequently, the substrates were cooled down to $480^{\circ}$C under As$_4$ flux for the growth of the GaSb buffer layer with an Sb/Ga beam equivalent pressure (BEP) ratio of 7. We note that the both the GaAs and GaSb layers are highly insulating, as confirmed by control measurements. The substrates were then cooled down to 400 $^{\circ}$C under Sb$_4$ flux, and further cooled down to 180 $^{\circ}$C after closing the Sb shutter for the growth of Mn-doped Cd$_3$As$_2$. Once the sample temperature was stable at 180 $^{\circ}$C, we used the Cd, As, and Mn effusion cells with BEP of Cd around $6 \times 10^{-8}$  Torr, As around $2 \times 10^{-8}$  Torr and Mn around $1.2 \times 10^{-9}$  Torr.

\begin{figure*}
\includegraphics[width=0.9\textwidth]{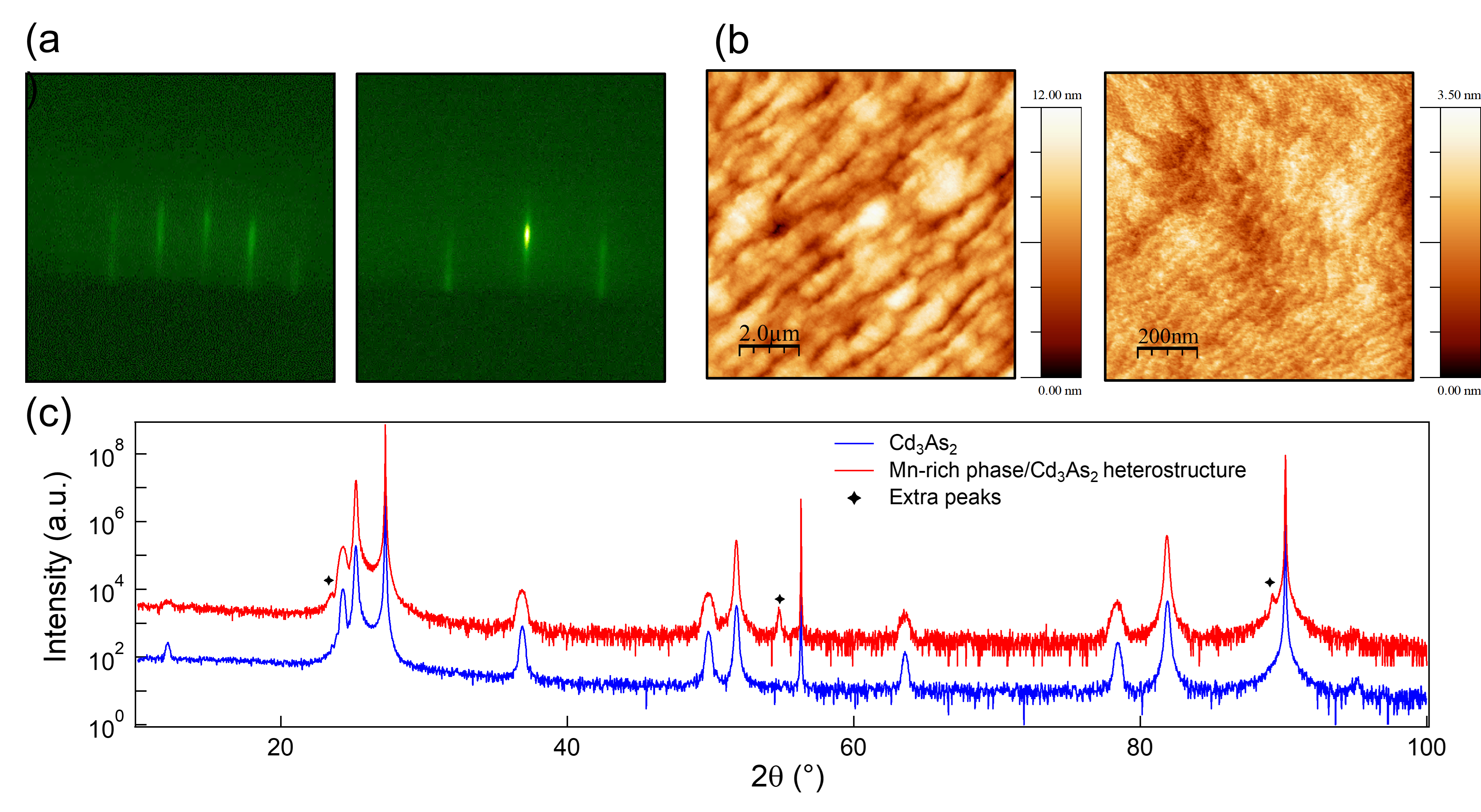}
\caption{\label{FIG.1} (a) {\it In situ} RHEED patterns during MBE of a nominally 15 nm thick Mn-doped Cd$_3$As$_2$ heterostructure grown using the uniform doping approach. The electron beam is directed along $[0\overline{1}1]$ (left) and $[\overline{2}11]$ (right) direction. (b) {\it Ex situ} AFM images of the same sample as in panel (a). (c) Out-of-plane XRD of a nominally 25 nm thick Mn-doped Cd$_3$As$_2$ heterostructure grown using the uniform Mn-doping approach. The ``extra peaks" in the XRD indicate the presence of a Mn-rich phase of unknown composition.
}
\end{figure*}

We attempted to grow Mn-doped Cd$_3$As$_2$ thin films using two methods: uniform doping wherein the Mn flux is constant during the growth of Cd$_3$As$_2$ and the delta doping method wherein we interrupt the growth of Cd$_3$As$_2$  and deposit a fractional monolayer of MnAs. The latter approach has been effective in magnetic doping of II-VI and III-V semiconductors \cite{crooker1995enhanced,Kawakami_APL_2000}. Both growth methods resulted in films with similar characteristics. Reflection high-energy electron diffraction (RHEED) measurements during the growth showed streaky patterns in either approach, indicating a relatively flat surface with some disorder even when depositing a Mn-rich layer (Fig. \ref{FIG.1}(a). Post-growth, we carried out {\it ex situ} AFM measurements (Fig. \ref{FIG.1}(b)) that indicated a root mean square surface roughness of 1.77 nm over an area of $10 \times 10 ~\mu \mathrm{m}^2$. The steps in the AFM image are due to the miscut substrate; these help prevent twinning defects and improve the quality of the sample. The $1 \times 1 ~\mu$m$^2$ AFM image in Fig. \ref{FIG.1}(b) shows the atomic steps of the heterostructure, indicative of the epitaxial growth of the heterostructure. Figure \ref{FIG.1}(c) compares the XRD scan of a nominally uniformly Mn-doped Cd$_3$As$_2$ film of nomila 25 nm thickness; the plot compares this XRD scan with that of a pristine Cd$_3$As$_2$ film. We observe extra diffraction peaks in the Mn-doped film suggesting the presence of an extra phase. However, from the XRD scan alone, we are unable to identify the crystal structure of this phase.   

To understand the crystalline structure and elemental distribution within the heterostructures, we used cross-sectional HAADF-STEM imaging and STEM-EDX elemental mapping. We prepared TEM lamella for STEM analysis on a FEI Helios Nanolab G4 dual-beam Focused Ion Beam (FIB). Amorphous carbon was first deposited on the films to protect the surface from damage due to exposure to the ion beam. STEM imaging and EDX spectroscopy were performed on an aberration–corrected FEI Titan G2 60-300 (S)TEM microscope equipped with a CEOS DCOR probe corrector and a super-X EDX spectrometer. The microscope was operated at 300 keV. We acquired HAADF-STEM images  with a probe convergence semi-angle of 25.5 mrad and detector inner and outer collection angles of 55 and 200 mrad respectively. We now discuss the TEM measurements obtained from a uniformly Mn-doped Cd$_3$As$_2$ film of nominal 15 nm thickness. Measurements taken on a delta-doped film of similar thickness yield qualitatively similar results (see Supplementary Material at xxx).  

\begin{figure*}
\includegraphics[width=0.9\textwidth]{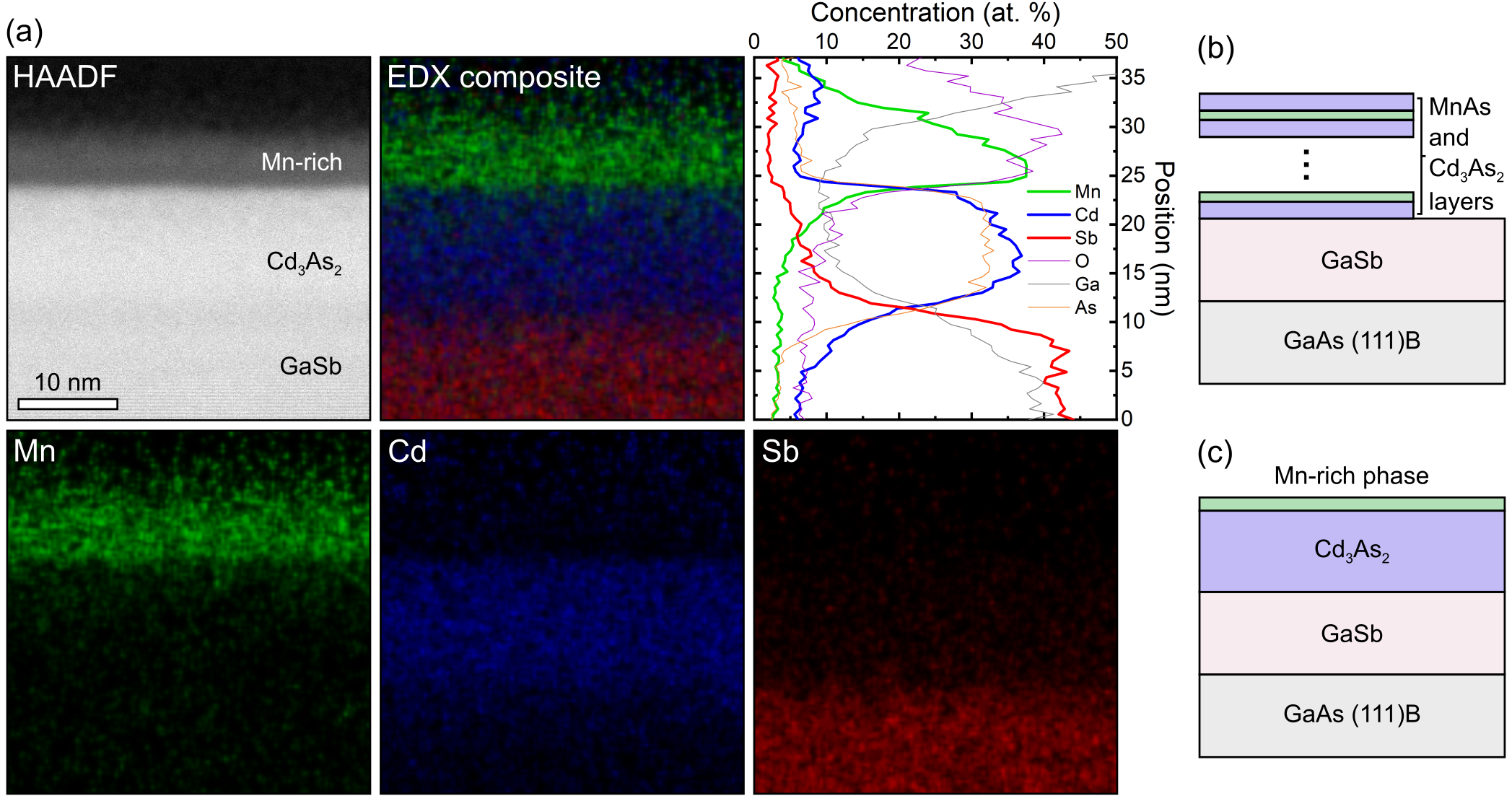}
\caption{\label{FIG.2} (a) Low-magnification cross-sectional HAADF-STEM image of a Mn-rich phase/Cd$_3$As$_2$ heterostructure that results from uniform Mn-doping during MBE growth of Cd$_3$As$_2$. Composite and individual EDX maps show the spatial distribution of Mn, Cd, and Sb in the heterostructure as well as an atomic percent concentration profile across the heterostructure. Instead of being incorporated into the Cd$_3$As$_2$ lattice, a segregated Mn-rich phase formed on the top of the Cd$_3$As$_2$ layer. (b) Schematic of the delta doping method. (c) Schematic of the actual Mn-rich phase/Cd$_3$As$_2$ heterostructure produced by the delta doping growth.}
\end{figure*}

High magnification HAADF-STEM images of the film cross-section revealed a crystalline Mn-Cd$_3$As$_2$ film of thickness $\sim 13$ nm, epitaxial to the GaSb substrate (Fig. \ref{FIG.2}(a)). A uniform amorphous-like layer ($\sim 5$ nm thick) of darker contrast is seen on top of the film. STEM-EDX elemental mapping, shown in Fig. \ref{FIG.2}(a), was used to obtain compositional information from the heterostructure. Surprisingly, instead of incorporating Mn throughout the film, the growth procedure, illustrated in Fig. \ref{FIG.2}(b), produced a segregated Mn-rich phase at the top surface of the Cd$_3$As$_2$ film. The resultant structure is visible in the EDX maps and illustrated in Fig. \ref{FIG.2}(c).

Averaging the results from several STEM-EDX data sets revealed that the darker contrast region seen in STEM images on top of the Cd$_3$As$_2$ film is primarily composed of a Mn-oxide phase with small traces of Cd and As (O: 49 at\%, Mn: 39 at\%; Cd and As: 12 at\%). While most of the Mn migrated to the surface, EDX analysis revealed $< 5$ at\% Mn content in the Cd$_3$As$_2$ film layer, indicating some doping of the films with Mn. 
In addition to the uniform Mn-rich layer, some regions exhibited islands rising out of the film surface with more mixing between the Mn, Cd and As (Cd: 17.3 at\%, Mn: 27.4 at\%, O: 24.5 at\%, As: 29.8 at\%). These islands were mostly amorphous with some crystalline regions closer to the surface of the film. The phase separation seen from the STEM-EDX data is supported by the low solubility of Mn in Cd$_3$As$_2$. As a result, even for the MBE growth under non-equilibrium conditions, it is hard to overcome the kinetic barrier to form Mn-doped Cd$_3$As$_2$\cite{ril2017phase}. 

Since our attempt to introduce magnetic dopants into Cd$_3$As$_2$ thin films inadvertently resulted in a phase-segregated heterostructure wherein a Mn-rich compound is cleanly interfaced with a Cd$_3$As$_2$ film, two important questions arise. First, is the Mn-rich phase magnetically ordered? If so, how does its presence affect the electronic transport properties of the Cd$_3$As$_2$ film with which it is directly interfaced? Naively, one might anticipate that interfacial exchange interaction between the magnetic moments in the overlayer and the band electrons in the Cd$_3$As$_2$ film would result in a degradation of the mobility due to spin-dependent scattering. We now address the magnetic and electrical transport properties of these Mn-doped Cd$_3$As$_2$ films. Since we have established that the actual sample structure consists of a thin Mn-rich overlayer interfaced with a Cd$_3$As$_2$ film regardless of the growth method used, we will refer to the samples being measured as heterostructures with $x$ nm Mn-rich layer/$y$ nm Cd$_3$As$_2$. 

\begin{figure*}
\includegraphics[width=0.9\textwidth]{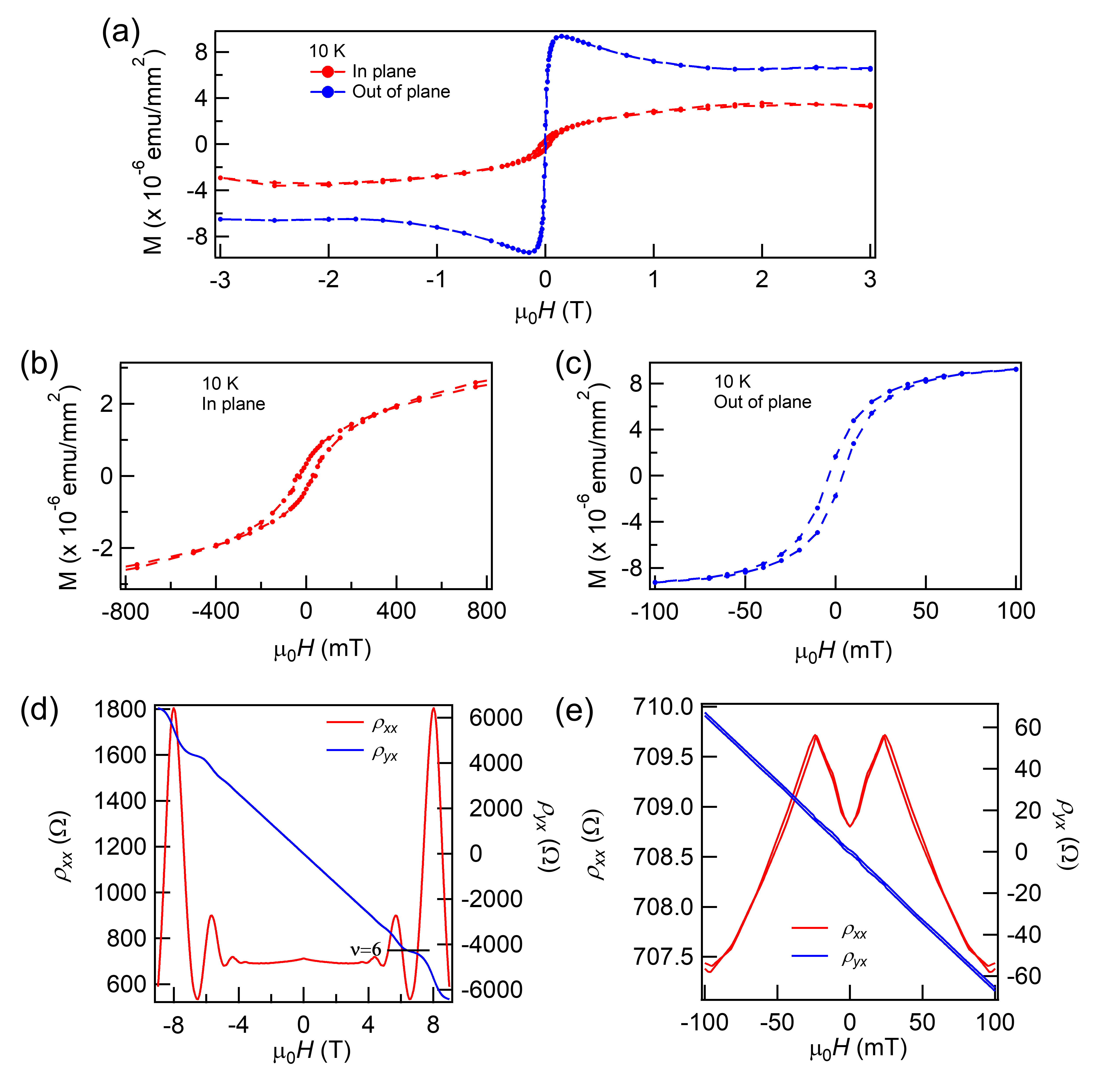}
\caption{\label{FIG.3}  Investigation of ferromagnetism in a Mn-rich layer/Cd$_3$As$_2$ heterostructure. (a), (b), (c) SQUID magnetometry measurements of a 1 nm Mn-rich phase/ 7 nm Cd$_3$As$_2$ heterostructure  at 10 K, respectively. The data suggest ferromagnetic ordering with an out-of-plane magnetic anisotropy. The inconsistency between the saturated magnetization values in the easy- and hard-axis measurements is an artifact created by the difficulty in properly handling the large filling factor when the thin film is mounted for field in plane measurements in the SQUID.   (d) Observation of Shubnikov de Haas oscillations and an incipient integer quantum Hall effect in a 1 nm Mn-rich phase/ 7 nm Cd$_3$As$_2$ heterostructure at 2 K. (e) An expanded view of the low field magnetoresistance and Hall effect in panel (d).  
}
\end{figure*}

We first discuss SQUID magnetometry measurements of a 1 nm Mn-rich layer/ 7 nm Cd$_3$As$_2$ heterostructure that results from the delta-doping method with 6 repeats of (sub-monolayer MnAs/1.25 nm Cd$_3$As$_2$); the thicknesses in the final structure are estimated from HAADF-STEM measurements of a thicker sample using similar conditions with twice the thickness. The measurements were carried out in a Quantum Design SQUID magnetometer with the sample mounted in a straw for field parallel to and field normal to the sample plane. Figure \ref{FIG.3} (a) shows the magnetization ($M$) versus field ($H$) at $T = 10$ K for this sample with the field in plane and out-of-plane. A diamagnetic background has been subtracted in these plots. Zoomed in views of $M$ vs. $H$ for field in-plane (Fig. 3(b)) and out-of-plane (Fig. 3(c)) show hysteresis loops with a very large coercive field ($\sim 80$ mT) in the former case and much smaller coercive field ($\sim 10$ mT) in the latter. The behavior of the coercive field and the saturation field for the two field orientations are consistent with ferromagnetism in the sample with a strong out-of-plane  magnetocrystalline anisotropy. This could arise from an inhomogenous distribution of Mn forming complex nanoscale cluster phases with the other elements present in the TEM analysis (primarily O, but also containing Cd and As). We note that the most commonly known compounds involving Mn and O (MnO and Mn$_3$O$_4$) are antiferromagnetic and feromagnetic, respectively. A more detailed understanding of the origin of ferromagnetism in such samples will require measurements such as x-ray magnetic circular dichroism and polarized neutron reflectometry.  

Next, we discuss electrical transport in this 1 nm Mn-rich layer/ 7 nm Cd$_3$As$_2$ heterostructure. We fabricated the samples into 40 $\mu \textrm{m} \times 10 ~\mu \textrm{m}$ Hall bar devices using photolithography and Ar$^+$ plasma dry etching. The top gate was defined by a 30 nm Al$_2$O$_3$ dielectric layer and Ti(5 nm)/Au(50 nm) contacts deposited by atomic layer deposition and electron beam evaporation, respectively.  Magnetoresistance (MR) and Hall effect measurements were carried out in a Quantum Design Physical Properties Measurement System over a temperature range 2 K $\leq T \leq 300$ K and in magnetic fields up to $B = 9$ T. All the MR and Hall effect data shown have been properly field symmetrized or anti-symmetrized, respectively. Figure 3(d) shows the results of such measurements at $T = 2$ K. Surprisingly, despite the presence of the ferromagnetic overlayer in direct contact with the Cd$_3$As$_2$ film, we observe pronounced quantum oscillations and even an incipient quantum Hall effect indicated by a Hall resistance plateau corresponding to $\rho_{xy} = \frac{1}{6}\frac{h}{e^2} = 4.3 ~\mathrm{k} \Omega$. We searched for possible signatures of an exchange coupling between the electrons in the Cd$_3$As$_2$ layer and Mn moments in the overlayer. Since the easy axis of the Mn-rich layer is out-of-plane according to our SQUID measurements, an exchange coupling between carriers in Cd$_3$As$_2$ and the Mn moments should lead to hysteresis in the field dependence of the Hall resistance and also to the MR from modifications to the quantum corrections to diffusive transport. Figure 3 (e) shows a careful sweep of the longitudinal MR and the Hall effect in the field range -0.1 T $\leq B \leq$ 0.1 T. The low field MR shows a non-monotonic dependence on magnetic field with an initial positive MR followed by a sudden change to a negative MR at a field close to the coercive field observed in SQUID magnetometry. The negative MR at fields higher than the saturation field of the ferromagnetic layer may indicate a reduction in spin-dependent scattering. In contrast, pristine Cd$_3$As$_2$ films grown under similar conditions only show a positive MR for field perpendicular to the sample plane (see Supplementary material at xxx). We do not see any obvious signs of hysteretic or non-linear Hall resistance. It is possible that with the step size used in the magneto-transport measurements (0.01 T), we might not be able to resolve these differences given how quickly the magnetization saturates with field along the anisotropy axis and the narrowness of hysteresis loop.

\begin{figure*}
\includegraphics[width=0.9\textwidth]{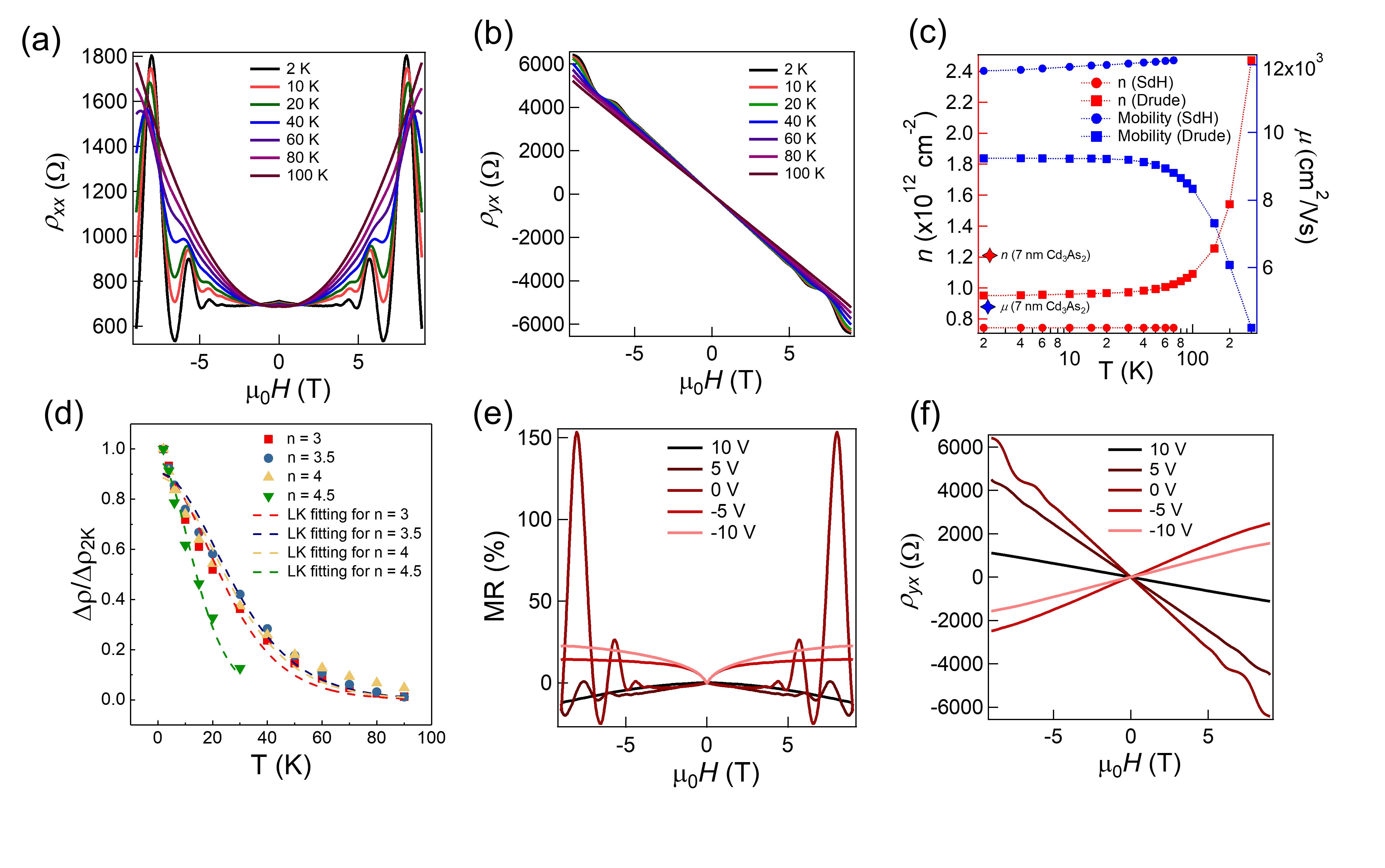}
\caption{\label{FIG.4} Quantum oscillations in a 1 nm Mn-rich phase/ 7 nm Cd$_3$As$_2$ heterostructure. (a), (b) The temperature dependence of the quantum oscillations and Hall effect, respectively. (c) Temperature dependence of the carrier density and mobility. The values of the carrier density and mobility for a pristine Cd$_3$As$_2$ film of the same thickness are also shown at 2 K for comparison. (d) The fitting for the amplitude of the quantum oscillations at different filling factors with the Lifshitz-Kosevich formula. (e), (f) Gate dependence of the quantum oscillations and Hall effect, respectively.
}
\end{figure*}

To further understand the effect of the Mn-rich overlayer on quantum transport, we analyzed the quantum oscillations in the heterostructure and compared the behavior with similar measurements in pure Cd$_3$As$_2$ thin films. For the 1 nm Mn-rich layer/ 7 nm Cd$_3$As$_2$ heterostructure, the amplitude of quantum oscillations gradually decreased with increasing temperature but remained finite up to about $T = 100$ K as shown in Fig.\ref{FIG.4}(a). We note that the $\nu = 6$ quantum Hall plateau mentioned earlier was observable up to $T = 20$ K in Fig. 4(b). To extract the carrier density, mobility, and effective mass of the carriers involved in quantum transport, we studied the temperature dependence of the quantum oscillations (Figs. \ref{FIG.4}(c) and \ref{FIG.4}(d)). The carrier density and mobility were calculated using both the Drude model and quantum oscillations. Using the Drude model, we find a carrier density $n = 9.5 \times 10^{11} \textrm{cm}^{-2}$ and a mobility $\mu = 9200$ cm$^{2}$/V/s at $T = 2$ K. The analysis of quantum oscillations yields $n = 7.2 \times 10^{11} \textrm{cm}^{-2}$ and $\mu = 12000$ cm$^{2}$/V/s. The higher carrier density and lower carrier mobility deduced from Drude analysis compared to that obtained from quantum oscillations indicates the presence of low-mobility carriers that do not contribute to the quantum oscillations. The band origin of the low mobility carriers is not clear at this stage. By fitting the amplitude of the quantum oscillations using the standard Lifshitz-Kosevich formula \cite{uchida2017quantum}, we extracted an effective mass $m^{*} = 0.04 m_e$. This light effective mass is due to the Dirac dispersion and is in good agreement with previous studies of Cd$_3$As$_2$ thin films \cite{zhao2015anisotropic,schumann2018observation,uchida2017quantum}.

Figure \ref{FIG.4}(c) also shows the carrier density and the mobility of a pure 7 nm Cd$_3$As$_2$ thin film (at $T = 2$K) calculated from the Drude model; this film was grown under nominally identical conditions (substrate temperature and Cd:As flux ratio) to those used for the growth of Mn-delta-doped samples. The carrier density in the pure film is about 20\% higher and the Drude mobility is about 50\% lower than in the Mn-doped film. Our findings suggest that the Mn-rich layer does not contribute to the transport signal and perhaps passivates the Cd$_3$As$_2$ layers underneath from electronegative OH-surface adsorbates \cite{galletti2018nitrogen}. 

Although the Mn-rich layer lies between the Cd$_3$As$_2$ layer and the top gate, it does not appear to affect the efficiency of the top gate. As shown in Fig. \ref{FIG.4}(e) and \ref{FIG.4}(f), the chemical potential of the heterostructure can be tuned from n-type to p-type by applying a top gate voltage. The quantum oscillations are only observed in the positive gate voltage regime, indicating the higher mobility of n-type carriers. We also note that the emerging quantum Hall plateaus are observable only around zero gate voltage, which is close to the Dirac point of the Cd$_3$As$_2$ layer. Away from the Dirac point, the quantum Hall plateaus disappear because of the high carrier density in the n-type regime and the low mobility in the p-type regime. 

We now discuss our results in the context of prior studies of pristine (non-magnetic) Cd$_3$As$_2$ thin films, in particular those grown on (111) GaSb/GaAs. The integer quantum Hall effect was initially observed in a (112)-oriented Cd$_3$As$_2$ thin film of 20 nm thickness \cite{schumann2018observation}, followed by observations in similarly grown films whose thicknesses ranged from 10 nm - 60 nm \cite{Galletti_PhysRevB.99.201401}. While it is clear that the integer quantum Hall effect in these Cd$_3$As$_2$ films indicates quantum transport in a two-dimensional (2D) electron gas, a complete understanding of its origin is still lacking. A systematic study of the thickness-dependence of the quantum transport appears to rule out the role of Weyl orbits in the experimentally observed quantum Hall effect. \cite{Galletti_PhysRevB.99.201401} This hypothesis is consistent with our observation of an incipient integer quantum Hall effect in an even thinner (7 nm) Cd$_3$As$_2$ film than in these earlier studies; in addition, we observe the quantum Hall effect in the presence of a highly structurally disordered ferromagnetic interface at the top surface. We note that recent density functional theory calculations indicate that surface Fermi arcs likely survive in Cd$_3$As$_2$ films down to the thicknesses studied experimentally here. \cite{Arribi_PhysRevB.102.155141} However, these Fermi arc states on the top surface would presumably not survive in the presence of the disordered magnetic layer, especially with an out-of-plane magnetization. Our results appear to suggest that the wave function of the 2D electron system in ultrathin Cd$_3$As$_2$ films, likely built from Fermi arc states and quantum confined bulk states, has a maximum located away from the top and bottom surfaces, thus being immune to perturbation by the magnetism on the top surface. At this stage, however, we do not have a rigorous calculation to support this speculation.      

\section{Summary}
In summary, we have shown that conventional magnetic doping approaches to introducing ferromagnetism into Cd$_3$As$_2$ by MBE result in a phase separation with a Mn-rich near-surface layer capping a pristine Cd$_3$As$_2$ film since Mn likely acts as surfactant. The Mn-rich region shows out-of-plane magnetic anisotropy in SQUID measurements. Surprisingly, this ferromagnetic Mn-rich layer does not negatively affect the quantum transport in the Cd$_3$As$_2$ underneath. On the contrary,  compared with pure Cd$_3$As$_2$ thin films grown under identical conditions, the presence of the Mn surfactant lowers the carrier density so that the chemical potential is close to the Dirac point and leads to an enhanced electron mobility. The resulting 1 nm Mn-rich layer/ 7 nm Cd$_3$As$_2$ films show an emergent quantum Hall effect in transport measurements. This should become more robust if the samples are measured at lower temperature and at higher magnetic fields. Measurements at dilution fridge temperatures may also reveal emergent physics resulting from the presence of the ferromagnetic overlayer. Although it is hard to dope Mn directly into Cd$_3$As$_2$ using MBE, it is worthwhile using other techniques such as polarized-neutron reflectometry to search for magnetic proximity effects of the easy-plane ferromagnetism on the Cd$_3$As$_2$ underneath. Our study indicates that extreme caution is called for in relying on magnetotransport data alone as a sign of broken time-reversal symmetry in magnetically-doped Cd$_3$As$_2$ films, especially in the absence of electron microscopy data \cite{liu2018cr}.

\begin{acknowledgments}
This project was supported by the Institute for Quantum Matter under DOE EFRC grant DE-SC0019331 (RX,NS). The magnetometry measurements were supported by a grant from the University of Chicago (JR,NS). The electron microscopy effort (JH,AM) was supported by SMART, one of seven centers of nCORE, a Semiconductor Research Corporation program, sponsored by the National Institute of Standards and Technology (NIST) and by the College of Science and Engineering Characterization Facility, University of Minnesota, which has received capital equipment funding from the National Science Foundation through the UMN MRSEC under Award Number DMR-2011401. We thank Chris Leighton, Paul Crowell, Yi Li, and Junyi Zhang for useful comments.
\end{acknowledgments}

\nocite{*}

\providecommand{\noopsort}[1]{}\providecommand{\singleletter}[1]{#1}%

\newpage

\end{document}